\documentclass{mytlp}

\usepackage{graphics}
\usepackage[all]{xy}
\usepackage{times}
\usepackage{url}

\usepackage{amssymb}
\renewcommand{\tt}{\usefont{OT1}{cmtt}{m}{n}\selectfont}

\usepackage{wrapfig}

\def\boxit#1{\raisebox{-3.5pt}{\vbox{\hrule\hbox{\vrule\kern3pt
  \vbox{\kern3pt\hbox{#1}\kern3pt}\kern3pt\vrule}\hrule}}}

\newcommand{\inparen}[1]{\mbox{\rm (}#1\mbox{\rm )}}
\newcommand{\modulo}{\inparen{modulo a renaming of nodes}}

\newcommand{\equprogram}[1]{
\def\separator{1ex}
\frenchspacing
\refstepcounter{equation}%
\par\vspace\separator\hspace{2.5em}%
$\vcenter{\tt\noindent\kern-.5em{#1}}$%
\kern-35pt\llap{(\theequation)}
\par\vspace\separator\noindent\kern-.0em%
}
\newcommand{\code}[1]{\mbox{\tt #1}}   
\newcommand{\us}{\char95} 

\newcommand{\Roots}{\ensuremath{\mathcal{R}\mathrm{oots}}}
\newcommand{\nodes}{\ensuremath{\mathcal{N}}}

\newcommand{\xlabel}{\ensuremath{\mathcal{L}}}
\newcommand{\xsucc}{\ensuremath{\mathcal{S}}}

\newcommand{\vars}{\ensuremath{\mathcal{X}}}

\newcommand{\strat}{{\mathcal{S}}}

\newcommand{\COMMENT}[1]{}
\usepackage{color}
\newcommand{\todo}[1]{{\color{red}\sc [{#1}]}}

\usepackage{theorem}

\newtheorem{principle}{Principle}
\newtheorem{definition}{Definition}
\newtheorem{lemma}{Lemma}

\newtheorem{theorem}{Theorem}
\newtheorem{corollary}{Corollary}
\newcommand{\myproof}[1]{
  \ifx\showproof\undefined
  \else
    \mbox{\it Proof\ \ } #1 ~~$\Box$ \\[2ex]
  \fi
}
\def\showproof{}

\newcommand{\combstep}{\rlap{\raisebox{.2ex}[0pt][0pt]{\kern.05em$\sim$}}{\to}}

\def\vect(#1,#2){#1_1, \ldots, #1_{#2}}

\newcommand{\tostar}{\buildrel * \over \to}

\newcommand{\toplus}{\buildrel \mbox{\rm \tiny +} \over \to}
\newcommand{\pulltab}{\mbox{\raisebox{.15ex}[0pt][0pt]{\tiny$\Xi$}}}
\newcommand{\toxi}{
  \mbox{$\rlap{\raisebox{.15ex}[0pt][0pt]{\kern.95ex\tiny$\Xi$}}\to\,$}
}
\newcommand{\toxistar}{\mbox{$\,\buildrel * \over \toxi\,$}}
\newcommand{\toequal}{\to^{\llap{\raisebox{.3ex}{\tiny $=$}}}}
\newcommand{\leftequal}{{}^{\rlap{\ \raisebox{.3ex}{\tiny $=$}}}\leftarrow}

\newcommand{\rep}{R}

\newcommand{\col}{{:}}

\newcommand{\TOY}{{\cal TOY}}

\begin{document}
\pagenumbering{arabic}
\setcounter{page}{713}


\title{On the Correctness of Pull-Tabbing}


\author[S. Antoy]
{
Sergio Antoy \\
  \vrule height3ex width0ex Computer Science Department \\
  Portland State University \\
  Portland, OR 97207, U.S.A. \\
  \vrule height3.5ex width0ex {\tt antoy@cs.pdx.edu}
}

\maketitle
\begin{abstract}
  Pull-tabbing is an evaluation approach for functional
  logic computations, based on a graph transformation recently proposed,
  which avoids making irrevocable non-deterministic
  choices that would jeopardize the completeness of computations.
  In contrast to other approaches with this property, it does not
  require an upfront cloning of a possibly large portion of the
  choice's context.
  We formally define the pull-tab transformation,
  characterize the class of programs for which the transformation
  is intended, extend the computations in these programs
  to include the transformation, and prove the correctness
  of the extended computations.
\end{abstract}


{%
 {\fontsize{6}{9pt}\sffamily
 \vbox to 0pt{
 \kern-4.5in
 \hbox {\emph{TLP} {\bf 11} (4--5): 713--730, 2011.
    \copyright\ Cambridge University Press 2011}
 \hbox {doi:10.1017/S1471068411000263}
 \hbox to 0pt {Draft Wed Mar 30 17:21:24 PDT 2011 \hss}
}}}%

\section{Introduction}

Functional logic programming \cite{AntoyHanus10CACM}
joins in a single paradigm the features of functional programming with those
of logic programming.  Logic programming contributes logic
variables that are seamlessly integrated in functional computations
by narrowing.  The usefulness and elegance of programming with narrowing
is presented in \cite{AntoyHanus02FLOPS,Antoy10JSC}.  
At the semantics level free variables are equivalent to
\emph{non-deterministic functions} \cite{AntoyHanus06ICLP},
i.e., functions that for some arguments may return any one of many results.
Thus, at the implementation level variables can be replaced
by non-deterministic functions when non-deterministic functions 
appear simpler, more convenient and/or more 
efficient to implement \cite{BrasselHuchAPLAS07}.
This paper focuses on a graph transformation recently proposed
for the implementation of non-determinism of this kind.
This transformation is intended to ensure
the completeness of computations without cloning too eagerly
a large portion of the context of a non-deterministic step.
The hope is that steps following the transformation will create conditions
that make cloning the not yet cloned portion of the context unnecessary.

\section{Motivation}

Non-determinism is certainly the most characterizing and appealing
feature of functional logic programming.  It enables
encoding potentially difficult problems into relatively simpler programs.
For example, consider the problem of abstracting the
dependencies among the elements of a set such as
the functions of a program or the widgets of a graphical user interface.
In abstractions of this kind,
\emph{component parts} ``build'' \emph{composite objects}.
A non-deterministic function, \code{builds}, defines which
objects are dependent on each part.
The syntax is Curry~\cite{Hanus06Curry}.
\equprogram{
   builds p1 = o1 \\
   builds p1 = o2 \\
   builds p2 = o1 \\
   ...
}
A part can build many objects, e.g.:
part \code{p1} builds objects \code{o1} and \code{o2}.
Likewise, an object can be built from several parts,
e.g.: object \code{o1} is built by parts \code{p1} and \code{p2}.
Many-to-many relationships, such as that between
objects and parts just sketched,
are difficult to abstract and to manipulate in deterministic languages.
However, in a functional logic setting, the non-deterministic
function \code{builds} is straightforward to define and
is sufficient for all other basic functions of the abstraction.

For example, a function that non-deterministically computes a part
of an object is simply defined by:
\equprogram{
   is\_built\_by (builds x) = x
}
where \code{is\_built\_by} is defined using 
a \emph{functional pattern} \cite{AntoyHanus05LOPSTR}.
The set of all the parts of an object is computed by
\code{is\_built\_by'set}, the implicitly defined \emph{set function}
\cite{AntoyHanus09PPDP} of \code{is\_built\_by}.

The simplicity of design and ease of coding offered by
functional logic languages through non-determinism do not come for free.
The burden unloaded from the programmer is placed on the execution.
All the alternatives of a non-deterministic
choice must be explored to some degree to ensure that no result of a
computation goes missing.
Doing this efficiently is a subject of active research.
Below, we summarize the state of the art.

\section{Approaches}
\label{Approaches}

There are three main approaches to the execution of
non-deterministic steps in a functional logic program.
A fourth approach, called \emph{pull-tabbing} \cite{Alqaddoumi10GCM},
still underdeveloped, is the subject of this paper.  
Pull-tabbing offers some appealing
characteristics missing from the other approaches.

\subsection{Running example}
\label{Running example}

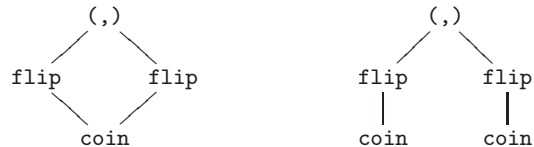
\begin{figure}[ht]
  \centering
  \begin{minipage}{1in}
    \xymatrix@1@C=5pt@R=12pt@M=1pt{
      & \code{(,)} \ar@{-}[dl] \ar@{-}[dr] \\
      \mathtt{flip} \ar@{-}[dr] & & \mathtt{flip} \ar@{-}[dl] \\
      & \mathtt{coin} 
    }
  \end{minipage}
  \hspace*{6em}
  \begin{minipage}{1in}
    \xymatrix@1@C=5pt@R=12pt@M=1pt{
      & \code{(,)} \ar@{-}[dl] \ar@{-}[dr] \\
      \mathtt{flip} \ar@{-}[d] & & \mathtt{flip} \ar@{-}[d] \\
      \mathtt{coin} & & \mathtt{coin}
    }
  \end{minipage}
  \caption{\it Depiction of \emph{(\protect{\ref{value}})}
    \inparen{left} and of \code{(flip coin, flip coin)}
    \inparen{right} as graphs. The symbol \code{(,)} denotes the
    pair constructor.}
  \label{fig:sharing}
\end{figure}

We borrow from \cite{Alqaddoumi10GCM} a simple example
to present the existing approaches and understand
their characteristics:
\equprogram{
flip 0 = 1 \\
flip 1 = 0 \\
coin = 0 ? 1
}
We want to evaluate the expression
\equprogram{
\label{value}
(flip x, flip x) where x = coin
}
We recall that `\code{?}' is a library function, called
\emph{choice}, that returns either of its arguments,
i.e., it is defined by the rules:
\equprogram{
\label{binary-choice-rules}
x ? \us{} = x \hspace*{8em}{\rm rule} $C_1$ \\
\us{} ? y = y \hspace*{8em}{\rm rule} $C_2$
}
and that the \code{where} clause introduces
a \emph{shared} expression.  Every occurrence of \code{x} in
(\ref{value}) has the same value throughout the entire computation
according to the \emph{call-time choice} 
semantics~\cite{Hussmann92JLP,Lopez-FraguasRodriguez-HortalaSanchez-Hernandez07PPDP}.
By contrast, in \code{(flip\,coin,\,flip\,coin)} each occurrence of
\code{coin} is evaluated independently of the other.~
Fig.~\ref{fig:sharing} highlights the difference between these two 
expressions when they are depicted as graphs.

\noindent
A \emph{context} is an expression with a
distinguished symbol called \emph{hole} denoted `$[\,]$'.
If $C$ is a context, $C[x]$ is the expression obtained by replacing the
hole in $C$ with $x$.  E.g., the expression in (\ref{value})
can be written as $C[\code{coin}]$, in which
$C$ is \code{(flip x, flip x) where x = $[\,]$}.
The context $[\,]$ is called \emph{empty} context.
An expression rooted by a node $n$ labeled by the choice symbol
is informally referred to as \emph{a choice}
and each argument of the choice symbol, or successor of $n$,
is referred to as a choice's \emph{alternative}.

\subsection{Previous approaches}

\emph{Backtracking} is the most traditional approach to
non-deterministic computations in functional logic programming.
Evaluating a
choice in some context, say $C[u?v]$, consists in selecting
either alternative of the choice, e.g., $u$ (the criterion
for selecting the alternative is not relevant to our discussion),
replacing the choice with the selected alternative, which gives
$C[u]$, and continuing the computation.
In typical interpreters, if and when the computation of $C[u]$ completes,
the result is consumed, e.g., printed,
and the user is given the option to either terminate the execution
or compute $C[v]$.
Backtracking is well-understood and relatively simple to implement.
It is employed in successful languages such as Prolog
\cite{IsoProlog} and in language implementations such as
PAKCS~\cite{Hanus08PAKCS} and  $\TOY$ \cite{ToyHomepage}.
The major objection to backtracking is its incompleteness.
If the computation of $C[u]$ does not terminate,
no result of $C[v]$ is ever obtained.
\\[2ex]
\emph{Copying} (or \emph{cloning}) is an approach that
fixes the inherent incompleteness
of backtracking.  Evaluating a choice in some context, say
$C[u?v]$, consists in evaluating simultaneously (e.g., by
interleaving steps) and independently both $C[u]$ and $C[v]$.  In
typical interpreters, if and when the computation of either
completes, the result is consumed, e.g., printed, and the user is
given the option to either terminate the execution or continue with the
computation of the other.  Copying is
simpler than backtracking and it is used in some experimental
implementations of functional logic languages
\cite{AntoyHanusLiuTolmach04IFL,TolmachAntoyNita04icfp}.
The major objection to copying is the significant investment
of time and memory made when a non-deterministic step is executed.
If an alternative of a choice eventually fails, cloning the context
may have been largely useless.  For a contrived example, notice
that in \code{1+(2+($\ldots$+($n$ `div` coin)$\ldots$))}
an arbitrarily large context is cloned when the choice is
evaluated, but for one alternative this context
is almost immediately discarded.
\\[2ex]
\emph{Bubbling} is an approach proposed
to avoid the drawbacks of backtracking and copying 
\cite{AntoyBrownChiang06Termgraph,Lopez-Fraguas08FLOPS}.
Bubbling is similar to copying, in that it clones a portion of the
context of a choice to concurrently compute all its alternatives,
but the portion of cloned context is typically smaller
than the entire context.
We recall that
in a rooted graph $g$, a node $d$ is a \emph{dominator} of a node
$n$, \emph{proper} when $d\ne n$, iff every path from the root of $g$ to $n$
contains $d$.
An expression $C[u?v]$ can be seen as $C_1[C_2[u?v]]$ in which
the root of $C_2$ is a dominator of the hole.  A trivial case
arises when $C_1=[\,]$ and $C_2=C$.
Evaluating a choice in some context, say
$C[u?v]$, distinguishes whether or not $C$ is empty.
If $C$ is the empty context, $u$ and $v$ are evaluated
simultaneously and independently, as in copying,
but there is no context to clone.
Otherwise, the evaluation consists in finding $C_1$ and $C_2$ such that
$C[u?v]=C_1[C_2[u?v]]$ and the root of $C_2$ is a proper dominator
of the choice, and evaluating $C_1[C_2[u]?C_2[v]]$.
When $C_1$ is the empty context, then bubbling is exactly
as copying.  Otherwise a smaller context, i.e., $C_2$ instead of
$C$, is cloned.
Bubbling intends to reduce cloning in hopes that some alternative
of a choice will quickly fail.
\begin{wrapfigure}[12]{r}{2.45in}
  \vspace*{-1ex}
  \begin{center}
    \boxit{
    \xymatrix@1@C=5pt@R=12pt@M=1pt{
      & & & \code{?} \ar@{-}[dll] \ar@{-}[drr] \\
      & \code{(,)} \ar@{-}[dl] \ar@{-}[dr] & & & & \code{(,)} \ar@{-}[dl]
      \ar@{-}[dr] \\
      \mathtt{flip}  \ar@{-}[dr] &  & \mathtt{flip} \ar@{-}[dl] &  & \mathtt{flip}
      \ar@{-}[dr] &  & \mathtt{flip} \ar@{-}[dl]  \\
      & \mathtt{0} & & & &\mathtt{1}
    }
  }
  \end{center}
  \vspace*{1ex}
 ~\begin{minipage}{2.4in}
  \caption{\it Graph depiction of the state of the computation
    of \emph{(\protect{\ref{value}})} after a bubbling step.
    Since the dominator of the choice is the root, bubbling
    and copying are the same for this example.}
  \label{fig:bubbling}
  \end{minipage}
\end{wrapfigure}
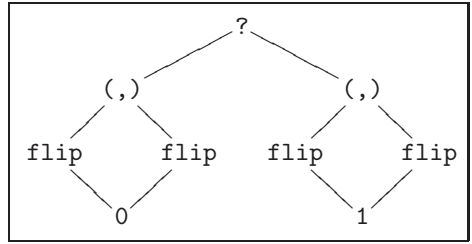

\noindent
An objection to bubbling is the cost of finding
a choice's immediate dominator and the risk of
paying this cost repeatedly for the same choice.
This cost entails traversing a possibly-large portion of the
choice's context.
Traversing the context is more efficient than cloning it, since
cloning requires node construction in addition to the traversal,
but it is still unappealing, since the cost of a non-deterministic
step is not predictable and it may grow with the size of an expression.

\subsection{Pull-Tabbing}

\emph{Pull-tabbing}, which is at the core of our work, 
was first sketched in \cite{Alqaddoumi10GCM}.
The name ``pull-tab'' originates from the metaphor
of pulling the tab of a zipper.  For an expression,
a choice is a tab and a choice's spine is a zipper.
As the tab/choice is pulled up, the zipper/spine opens
into two equal strands each of which has a different
alternative of the choice at the end.

Evaluating a choice in some context, say $C[u?v]$,
distinguishes whether or not $C$ is empty.
If $C$ is empty, $u$ and $v$ are evaluated
simultaneously and independently, as in copying
and bubbling, without any context to clone.
Otherwise, the expression to evaluate is of the form
$C[s(u?v)]$, for some symbol $s$ (for ease of
presentation we assume that $s$ is unary, but there are no restrictions
on its arity) and some context $C$.
Pull-tabbing transforms the expression into $C[s(u)?s(v)]$.
Without some caution, this transformation is unsound.

\begin{figure}[ht]
  \centering
  \resizebox{.98\textwidth}{!}{
  \begin{minipage}{1in}
    \xymatrix@1@C=5pt@R=10pt@M=1pt{
      & \code{(,)} \ar@{-}[dl] \ar@{-}[dr] \\
      \mathtt{flip} \ar@{-}[dr] & & \mathtt{flip} \ar@{-}[dl] \\
      & \mathtt{coin} 
    }
  \end{minipage}
  \hspace*{2ex} $\to$ \hspace*{1ex} 
  \begin{minipage}{1in}
    \xymatrix@1@C=5pt@R=10pt@M=1pt{
      & \code{(,)} \ar@{-}[dl] \ar@{-}[dr] \\
      \mathtt{flip} \ar@{-}[dr] & & \mathtt{flip} \ar@{-}[dl] \\
      & \mathtt{\code{?}} \ar@{-}[dr] \ar@{-}[dl] \\
      \mathtt{0} & & \mathtt{1}
    }
  \end{minipage}
  \hspace*{2ex} $\to$ \hspace*{1ex} 
  \begin{minipage}{1in}
    \xymatrix@1@C=5pt@R=10pt@M=1pt{
      & & \code{(,)} \ar@{-}[dl] \ar@{-}[dr] \\
      & \code{?} \ar@{-}[dl] \ar@{-}[dr] & & \mathtt{flip} \ar@{-}[d] \\
      \mathtt{flip} \ar@{-}[dr] & & 
           \mathtt{flip}  \ar@{-}[dr] & \code{?}  \ar@{-}[dll] \ar@{-}[d]\\
      & \mathtt{0} & & \mathtt{1}
    }
  \end{minipage}
  \hspace*{2ex} $\to$ \hspace*{1ex} 
  \begin{minipage}{1in}
    \xymatrix@1@C=5pt@R=10pt@M=1pt{
      & & \code{(,)} \ar@{-}[dl] \ar@{-}[dr] \\
      & \code{?} \ar@{-}[dl] \ar@{-}[dr] & & \mathtt{flip} \ar@{-}[d] \\
      \mathtt{1} & & 
           \mathtt{0} & \code{?}  \ar@{-}[dll] \ar@{-}[d]\\
      & \mathtt{0} & & \mathtt{1}
    }
  \end{minipage}
  }
  \caption{\it Initial portion of the computation of 
    \emph{(\protect{\ref{value}})} with pull-tabbing.
    The choice in the second expression is being pulled up
    along the left path to the root.
    This computation would eventually produce several results
    including \code{(1,0)} which mixes the left and right 
    alternatives of the same choice.}
  \label{fig:pull-tab}
\end{figure}
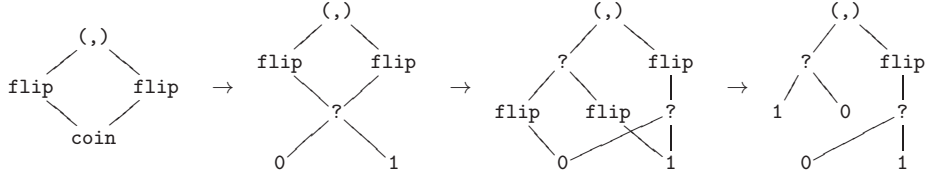

Unsoundness may occur when
some choice has two predecessors, as in our running example.
The choice will be pulled up along two paths creating 
\emph{two pairs} of strands
that eventually must be pair-wise combined together.
Some combinations will
contain mutually exclusive alternatives, i.e.,
subexpressions impossible to obtain in
languages such as Curry and $\TOY$ that adopt
the call-time choice semantics.
Fig.~\ref{fig:pull-tab} presents an example of this situation.

\noindent
We will show that the soundness is recovered
if the left and right
alternative of a choice are \emph{not} combined in the same expression.
To this aim, we attach an identifier to each
choice of an expression.
We preserve this identifier when a choice is pulled up.
If eventually the choice is reduced to either of its alternatives
every other choice with the same identifier
must be reduced to the same alternative.
A very similar idea in a rather different setting
is proposed in \cite{BrasselHuchAPLAS07,Brassel2011PhD}.

A pull-tab step clones a single node, 
a predecessor of the choice being pulled up.
If the choice is pulled all the way up to the root of an expression,
the choice's entire spine is cloned.
But if an alternative of the choice fails before the choice
reaches the root,
further cloning of the choice's context becomes unnecessary. 

\section{Formalization}

\subsection{Background}

We define a term graph in the customary
way~\cite{EchahedJanodet97IMAG},
but extend the decorations of nodes
with choice identifiers.
\begin{definition}[Expression]
\label{def:expression}
Let $\Sigma$ be a \emph{signature},
$\vars$ a countable set of \emph{variables},
$\nodes$ a countable set of \emph{nodes},
$\Omega$ a countable set of \emph{choice identifiers}.
A \emph{(rooted)} \emph{graph} over
$\langle\Sigma,\nodes,\vars,\Omega\rangle$
is a 5-tuple 
$g=\langle \nodes_g,\xlabel_g,\xsucc_g,$ $\Roots_g,id_g\rangle$
such that: 
\begin{enumerate}
\renewcommand{\labelenumi}{\arabic{enumi}.}
\item {}
$\nodes_g \subset \nodes$ is the set of nodes of $g$;
\item {}
$\xlabel_g : \nodes_g \to \Sigma \cup \vars$ is the \emph{labeling}
function mapping each node of $g$ to a signature symbol or a variable;
\item {}
$\xsucc_g : \nodes_g \to \nodes_g^*$
is the \emph{successor}
function mapping each node of $g$ to a possibly empty string of
nodes of $g$ such that
if $\xlabel_g(n)=s$, where $s \in \Sigma \cup \vars$,
and (for the following condition, we assume that a variable has arity zero)
$\mathit{arity}(s)=k$,
then there exist $n_1,\ldots,n_k$ in $\nodes_g$ such that
$\xsucc_g(n)=n_1 \ldots n_k$;
\item{}
$\Roots_g \subseteq \nodes_g$ is a subset of nodes of $g$
called the \emph{roots} of $g$;
\item {} $id_g : \nodes_g \to \Omega$ is a partial function
mapping nodes labeled by the choice symbol to a choice identifier;
\item{}
if $\xlabel_g(n_1) \in \vars$ and $\xlabel_g(n_2) \in \vars$
and $\xlabel_g(n_1) = \xlabel_g(n_2)$, then $n_1 = n_2$,
i.e., every variable of $g$ labels one and only one node of $g$; and
\item {}
for each $n \in \nodes_g$, either $n \in \Roots_g$
or there is a path from $r$ to $n$
where $r \in \Roots_g$, i.e., every node of $g$ is reachable
from some root of $g$.
\end{enumerate}
A graph $g$ is called a \emph{term (graph)},
or more simply an \emph{expression}, if $\Roots_g$ is a singleton.
\end{definition}
Typically we will use ``expression'' when talking about programs
and ``graph'' when making formal claims.
Choice identifiers play a role in computations.
Thus, we will explicitly define the $id$ mapping
only after formally defining the notion of computation.
Term graphs can be seen, e.g.,
in Figs.~\ref{fig:sharing} and \ref{fig:bubbling}.
Every choice node of every graph of Fig.~\ref{fig:pull-tab}
would be decorated with the same choice identifier.
Choice identifiers are arbitrary and only compared for equality.
Node names are arbitrary and irrelevant to most purposes
and are typically omitted.
However, some definitions and proofs of our claims need to
explicitly refer to some nodes of a graph.
For this purpose, we adopt the \emph{linear notation for graphs}
\cite[Def.~4]{EchahedJanodet97IMAG}.
With this convention, the left graph of Fig.~\ref{fig:sharing}
is denoted \code{$n_0$:($n_1$:flip\ $n_2$:coin,\ $n_3$:flip\ $n_2$)},
where the node names are the italicized identifiers starting with `$n$'.
We also make the convention that names of nodes that do not need to be
referenced can be omitted, hence \code{(flip\ $n_2$:coin,\ flip\ $n_2$)}.
The latter is conceptually identical to (\ref{value}).
In the linear notation for graphs, infix operators are
applied in prefix notation, e.g., see Lemma~\ref{invariance-by-pull-tab}.
This practice eases understanding the correspondence between
a node identifier and the label of that node.

The definition of graph rewriting
\cite{EchahedJanodet97IMAG,Plump99Handbook}
is more laborious than, although conceptually very similar to,
that of term rewriting
\cite{BaaderNipkow98,Terese03,DershowitzJouannaud90handbook,Klop92Handbook}.
Sections 2 and 3 of \cite{EchahedJanodet97IMAG} 
formalize key concepts of graph rewriting such as
\emph{replacement}, \emph{matching}, \emph{homomorphism},
\emph{rewrite rule}, \emph{redex}, and \emph{step}
in a form ideal for our discussion.
Therefore, we adopt entirely these definitions, including their notations,
and only discuss the manipulation of choice identifiers,
since they are absent from \cite{EchahedJanodet97IMAG}.


\subsection{Programs}

We now formalize the class of rewrite systems that we consider
in this paper.
A \emph{program} is a rewrite system in a class
called \emph{limited overlapping inductively sequential},
abbreviated \emph{LOIS}.
In \emph{LOIS} systems, the rules
are left-linear and constructor-based \cite{Odonnell85Book}.
The left-hand sides of the rules are organized
in a hierarchical structure called a 
\emph{definitional tree} \cite{Antoy92ALP}
that guides the evaluation strategy \cite{Antoy05JSC}.
In \emph{LOIS} systems, there is a single
operation whose rules' left-hand sides overlap.
This operation is called \emph{choice},
is denoted by the infix binary operation ``\code{?}'',
and is defined by the rules of (\ref{binary-choice-rules}).

\emph{LOIS} systems have been investigated in some depth.
Below we highlight informally the key results that
justify our choice of \emph{LOIS} systems.
\begin{enumerate}
\item  Any \emph{LOIS} system admits a complete, sound and
  optimal evaluation strategy \cite{Antoy97ALP}.
\item Any constructor-based conditional rewrite system is
  semantically equivalent to a \emph{LOIS} system \cite{Antoy01PPDP}.
\item Any \emph{narrowing} computation in a \emph{LOIS} system 
  is semantically equivalent to a \emph{rewriting} computation
  in another similar \emph{LOIS} system \cite{AntoyHanus06ICLP}.
\end{enumerate}
For the above reasons, \emph{LOIS} systems are an ideal core
language for functional logic programs.  Informally summarizing,
\emph{LOIS} systems are general enough to perform any functional
logic computation \cite{Antoy01PPDP} and powerful enough to
compute by simple rewriting \cite{AntoyHanus06ICLP} and without
wasting steps \cite{Antoy97ALP}.

\subsection{Computations}
\label{Computations}

In our setting, a \emph{computation} of $e$ is a sequence 
$e=e_0 \toxi e_1 \toxi \ldots$ such that $e_i \toxi e_{i+1}$ is a
\emph{step}, i.e., is either one of two graph transformations: a
rewrite, denoted by ``$\to$'', or a pull-tab, denoted by
``$\pulltab$''.
A \emph{rewrite} is the replacement in a graph of an instance
of a rewrite rule left-hand side (the \emph{redex}) with the corresponding
instance of the rule right-hand side (the \emph{replacement}).
The pull-tab transformation is formally defined in the next section.
In principle, we do not exclude choice reductions, i.e.,
non-deterministic steps, but in practice
we limit them to the root of an expression.
The reason is that reducing a choice makes an irrevocable commitment to one
of its alternatives.
Pull-tab steps are equivalent to non-deterministic steps in the
sense, formally stated and proved in the next section, that they
produce all and only the same results, but without
any irrevocable commitment.

A computation can be finite or infinite.
A computation is \emph{successful} or it \emph{succeeds} iff its
last element is a \emph{value}, i.e., a constructor normal form.
A computation is a \emph{failure} or it \emph{fails} iff its
last element is a normal form with some node labeled by an operation symbol.
In non-deterministic programs, such as those considered in
this paper, the same expression may have both successful
computations and failures.
Each expression of a computation is also referred to as
a \emph{state} of the computation.

A strategy determines which step(s) of an expression to execute.
Essential properties of a strategy, such as to succeed whenever
possible, will be recalled in Sec.~\ref{sec:correctness}.

\subsection{Transformations}

As described in Section \ref{Computations}, a computation
is a sequence of expressions such that each expression
of the sequence, except the first one, 
is obtained from the preceding expression by either of two transformations.
One transformation is an ordinary \emph{redex replacement}.
We defer to \cite[Def. 23]{EchahedJanodet98JICSLP} the 
precise formulation of this transformation and
to the next section the handling of decorations by this transformation.
The second transformation is defined below.
\begin{definition}[Pull-tab]
\label{def:pull-tab}
Let $g$ be an expression, $n$ a node of $g$,
referred to as the \emph{target},
not labeled by the choice symbol and 
$s_1\ldots s_k$ the successors of $n$ in $g$.
Let $i$ be an index in $\{1,\ldots k\}$ such that $s_i$,
referred to as the \emph{source},
is labeled by the choice symbol and
let $t_1$ and $t_2$ be the successors of $s_i$ in $g$.
Let $g_j$, for $j=1,2$, be the graph whose root
is a fresh node $n_j$ with the same label as $n$
and successors $s_1\ldots s_{i-1} t_j s_{i+1} \ldots s_k$.
Let $g'=g_1 \code{?} g_2$.
The \emph{pull-tab} of $g$ with source $s_i$ and target $n$ is
$g[n \leftarrow g']$ and we write $g \,\pulltab\, g[n \leftarrow g']$.
\end{definition}
Fig.~\ref{fig:pull-tab} depicts the result of a pull-tab step.
For a trivial textual example,
$((0+2)\,?\,(1+2))*3$ is the pull-tab of
$((0\, ?\, 1) + 2) * 3$.
The definition excludes
targets labeled by the choice symbol.  
These targets are not a problem
for the pull-tab transformation, 
but would complicate, without any benefit,
our treatment.

A pull-tab step is conceptually very similar to
an ordinary step---in a graph a (sub)graph is replaced.
The difference with respect to a rewrite step is that
the replacement is not constructed by a rewrite rule,
but according to Def.~\ref{def:pull-tab}.
It seems very natural for pull-tab steps too
to call \emph{redex} the (sub)graph being replaced.

Term and graph rewriting are similar formalisms
that for many problems are able
to model the expressions manipulated by functional logic programs.
Not surprisingly, expressions are terms in term rewriting
and graphs in graph rewriting.
A significant difference between these formalisms is
the identification of a subexpression of an expression.  
Term rewriting uses positions, i.e., paths in a tree, whereas graph
rewriting uses nodes.  Nodes are used for defining
both rewrite rules and expressions to evaluate.
%
\COMMENT{"In service" is a tax term used to identify when a
  depreciable asset is ready and available for a specific use
  then, the "placed in service" date is when the asset (equipment,
  real estate, vehicle, etc.) is eligible to be depreciated.}
Nodes are ``placed in service'' (1) to define rewrite rules,
(2) when an expression, called \emph{top-level},
is defined or created for the purpose of a computation, 
and (3) to define or construct the replacement
used in a step.  We agree that any node is placed in service
\emph{only once},  i.e., the same node is never
allocated to distinct top-level expressions and/or replacements.
However, the same node may be found in distinct graphs related by a 
step, since a step makes a localized change in a graph.
These stipulations are formalized by the following principle,
which is a consequence of placing nodes in service
\emph{only once}.

\begin{principle}[Persistence]
\label{persistence}
Let $g_1$ and $g_2$ be graphs.
If $n$ is a node in $\nodes_{g_1} \cap\, \nodes_{g_2}$,
then there exists a graph $g$ such that 
$g \toxistar g_1$ and $g \toxistar g_2$.
\end{principle}

\subsection{Decorations}

To support pull-tabbing and ensure its correctness
we attach additional information to an expression.
This additional information is formalized as a decoration
of a node similar to other decorations present in graph,
e.g., label and successors.
In this section, we rigorously define the function
that maps nodes to choice identifiers.

\begin{definition}[Decorations]
\label{decorations}
Let $A: g_0 \toxi g_1 \toxi \ldots$ be a computation.
We define the $id_{g_i}$ mapping,
for each element $g_i$ of $A$,
by induction on $i$, as follows:
$id_{g_i}$ takes a node of $g_i$ labeled by the choice symbol
and produces the node's choice identifier.
Base case: $i=0$.  
$id_{g_0}(n)$, where $n$ is in $g_0$ and is labeled by the choice symbol,
is an arbitrary element of $\Omega$, provided that $id_{g_0}$ is one-to-one.
Ind.~case: $i>0$.  By the induction hypothesis, 
$id_{g_{i-1}}$ is defined for any choice node.
In the step $g_{i-1} \toxi g_i$, whether rewrite or pull-tab,
a subexpression of $g_{i-1}$
rooted by a node $p$ is replaced by an expression rooted by a node $q$. 
Let $n$ be a node of $g_i$.
\begin{enumerate}
\setlength{\itemsep}{-5pt}
\item If $n$ is a node of $g_{i-1}$ labeled by the choice symbol,
then $id_{g_i}(n) = id_{g_{i-1}}(n)$. \\[-1ex]
\item Otherwise, if $g_{i-1} \to g_i$ \inparen{an ordinary rewrite}
  and $n$ is labeled by the choice symbol,
  then $id_{g_i}(n) = \alpha$, for an arbitrary $\alpha \in \Omega$
  provided that $id_{g_j}(m) \ne \alpha$
  for all $j < i$ and all $m \in g_j$ and $id_{g_i}(n) \ne id_{g_i}(m)$
  for $n \ne m$ \inparen{i.e., $\alpha$ is fresh}. \\[-1ex]
\item Otherwise, if $g_{i-1} \,\pulltab\, g_i$ 
  \inparen{a pull-tab} and $n=q$,
  then $id_{g_i}(n)=id_{g_{i-1}}(m)$, 
  where $m$ is the source node of the pull-tab.
\end{enumerate}
\end{definition}
The above definition is articulated, but conceptually simple.
Below, we give an informal account of it.
In a typical step $g \toxi g'$, most nodes of $g$ end up in $g'$.
The choice identifier, for choices,
of these nodes remains the same.
In a rewrite, some nodes are created.
Any choice node created in the step
gets a fresh choice identifier.
In a pull-tab, informally speaking,
the source (a choice)  ``moves'' and the target (not a choice) ``splits.''
The choice identifier ``moves'' with its source.
Split nodes have no choice identifier.

Each node in the ``universe'' of nodes $\nodes$ may belong to
several graphs.  In~\cite{EchahedJanodet97IMAG},
and accordingly in our extension (see Defs.~\ref{def:expression}
and \ref{decorations}),
the function mapping a node to a decoration
depends on each graph to which the node belongs.
It turns out that some decorations of a node,
e.g., the label, are \emph{immutable},
i.e., the function mapping a node to such decorations
does not depend on any graph.
We prove the immutability claim
for our extension, the  choice identifier.
Obviously, there is no notion of time when one discusses
expressions and considers the decorations of a node.
Hence immutable decorations ``are set'' with the nodes.
In practice, these decorations ``become known''
when a node is ``placed in service'' for the purpose of a computation
or is created by a step.

\def\lemmaimmut{
\begin{lemma}[Immutability]
\label{immutability}
Let $g_1$ and $g_2$ be expressions.
If $n$ is a node in $\nodes_{g_1} \cap \nodes_{g_2}$,
then $id_{g_1}(n)=id_{g_2}(n)$.
\end{lemma}
\myproof{
If a node $n$ belongs to $\nodes_{g_1} \cap \nodes_{g_2}$,
then, by Principle \ref{persistence},
there exists an expression $g$ and 
computations $A_1: g \toxistar g_1$ 
and $A_2: g \toxistar g_2$.
By induction on the length of $A_1$, resp.~$A_2$,
using point 1 of Def.~\ref{decorations},
$id_{g_1}(n)=id_{g}(n)$, resp. $id_{g_2}(n)=id_{g}(n)$.
The claim follows by transitivity.
}
}

\lemmaimmut
In view of this result, we drop 
the subscript from $id$ 
since this practice simplifies the notation
and attests a fundamental invariant.

Pull-tab steps may produce an expression
with distinct choices with the same choice identifier.
The same identifier tells us that to some extent these
redexes are the ``same''.  Therefore, when a computation
replaces one such redex with the left, resp.~right,
alternative, every other ``same'' redex should be replaced
with the left, resp.~right, alternative, too.
If this does not happen, the computation is unacceptable.
The notion of consistency of computations
introduced next abstracts this idea.

\begin{definition}[Consistency]
\label{def:consistent}
A rewrite step that replaces a redex rooted by a node $n$
labeled by the choice symbol is called a \emph{choice step}.
A computation $A$ is \emph{consistent}
iff for all $\alpha \in \Omega$, there exists an $i$
\mbox{\rm (\emph{either 1 or 2})} such that
every choice step of $A$ at a node identified by $\alpha$
applies rule $C_i$ of ``\,\code{?}'' defined in
\emph{(\ref{binary-choice-rules})}.
\end{definition}


\section{Correctness}
\label{sec:correctness}

A \emph{strategy} determines which step(s) of an expression to execute.
A strategy is usually defined as a function that takes
an expression $e$ and returns a set $S$ of steps of this expression or,
equivalently, the reducts of $e$ according to the steps of $S$.
We will not define any specific strategy.
A major contribution of our work is showing that the correctness
of pull-tabbing is strategy-independent.

The classic definition of correctness of a strategy $\strat$ is
stated as the ability to produce for any expression $e$ (in the
domain of the strategy) all and only the results that would be
produced by rewriting $e$.
``All and only'' leads to the following notions.
\\[1ex]
\emph{Soundness:} 
if $e \rlap{\raisebox{.2ex}{\kern1ex\tiny$\cal S$}}\tostar v$ is a
computation of $e$ in which each step is according to $\strat$
and $v$ is a value (constructor normal form), then
$e \tostar v$.  \\[1ex]
\emph{Completeness:} if $e \tostar v$,
where $v$ is a value (constructor normal form), 
then there exists a computation 
$e \rlap{\raisebox{.2ex}{\kern1ex\tiny$\cal S$}}\tostar v$
in which each step is according to $\strat$.
\\[1ex]
In the definitions of soundness and completeness proposed
above, the same expression is evaluated both
according to $\strat$ and by rewriting.
This is adequate with some conventions.
Rewriting is not concerned with choice identifiers.
This decoration can be simply ignored in rewriting computations.
In particular, in rewriting (as opposed to rewriting and pull-tabbing)
a computation is always consistent.
In graph rewriting, \emph{equality of graphs} is modulo a renaming of nodes.
A precise definition of this concept
is in \cite[Sect.~2.5]{EchahedJanodet97IMAG}.

Typically, the proof of soundness is trivial for strategies that
execute only rewrite steps, but our strategy executes also
pull-tab steps, hence it creates expressions that cannot be
produced by rewriting.  Indeed, some of these expressions
will have to be discarded to ensure the soundness.
The proof of correctness of pull-tabbing is
non-trivial and relies on two additional concepts,
\emph{representation} and \emph{invariance},
which are presented in following sections.

\subsection{Parallel Moves}

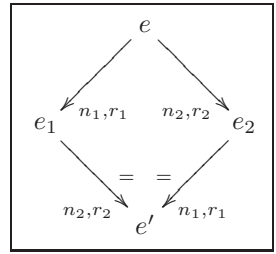
\begin{wrapfigure}[12]{r}{1.8in}
  \vspace*{-4ex}
  \begin{center}
    \boxit{
      \xymatrix@!C=3ex@!R=3ex{
        & e \ar@{->}[dl]^(.75){n_1,r_1}  \ar@{->}[dr]_(.75){n_2,r_2} \\
        e_1 \ar@{->}[dr]^(.65){=}_(.75){n_2,r_2} 
        & & e_2 \ar@{->}[dl]_(.65){=}^(.75){n_1,r_1}  \\
        & e'
      }
    }
  \end{center}
    \begin{minipage}{1.7in}
  \caption{\it The Parallel Moves Lemma
    for LOIS graph rewriting systems under appropriate conditions
    on nodes and rules.}
  \end{minipage}
  \label{fig:parallel-moves}
\end{wrapfigure}

Proofs of properties of a computation are often accomplished
by ``rearranging'' the computation's steps in some convenient order.
A fundamental result in rewriting, known as
the Parallel Moves Lemma \cite{HuetLevy91},
shows that in orthogonal systems the steps of a computation
can be rearranged at will.
A slightly weaker form of this result carries over
to \emph{LOIS} systems.
A pictorial representation of this result
is provided in Fig.~4. 
The symbol ``$\toequal$'' denotes the reflexive closure
of the rewrite relation.
The notation ``$\toequal_{n,r}$'', where $n$ is a node and $r$
is a rule, denotes either equality or a rewrite step at
node $n$ with rule $r$.

\def\lemmapmoves{
\begin{lemma}[\emph{LOIS} parallel moves]
\label{parallel-moves}
Let $e$, $e_1$ and $e_2$ be expressions such that
$e_1\ {_{n_1,r_1}}\kern-1.25ex\leftarrow e \to_{n_2,r_2} e_2$,
where for $i=1,2$, $n_i$ is a node and $r_i$ is a rule.
If $n_1 \ne n_2$ or both $n_1 = n_2$ and $r_1 = r_2$,
then there exists an expression $e'$ such that 
\modulo{}
$e_1 \toequal_{n_2,r_2}  e'\  {_{n_1,r_1}}\kern-1.25ex\leftequal e_2$.
\end{lemma}
\myproof{
By cases on the assumption's condition.
When both $n_1 = n_2$ and $r_1 = r_2$,
the two steps are the same, hence $e_1 = e' = e_2$.
When $n_1 \ne n_2$: the claim is a restriction
of \cite[Lemma 20]{Antoy97ALP} to rewriting in \emph{LOIS} systems.
}
}

\lemmapmoves
%

\subsection{Representation}

A characteristic of
pull-tabbing, similar to bubbling and copying,
is that the completeness of computations is obtained
by avoiding or delaying a commitment to
either alternative of a choice.
In pull-tabbing, similar to bubbling, \emph{both}
the alternatives of a choice are
kept or ``represented'' in a \emph{single} expression
throughout a good part of a computation.
The proof of the correctness of pull-tabbing
is obtained by reasoning about this concept,
which we formalize below.

\begin{definition}[Representation]
\label{representation}
We define a mapping $\rep$ that takes an expression $g$
and returns a set $\rep_g$ called the
\emph{represented set of $g$} as follows.
Let $g$ be an expression.
An expression $e$ is in $\rep_g$ iff
there exists a consistent computation $g \tostar e$
\modulo{}
that makes all and only the choice steps of $g$.
\end{definition}
In other words, we select either alternative for every
choice of an expression.  For choices with the same identifier,
we select the same alternative.
Since distinct choice steps occur at distinct nodes, by Lemma
\ref{parallel-moves} the order in which the choice steps are
executed to produce any member of the represented set 
is irrelevant.
Therefore, the notion of represented set is \emph{well defined}.
The notion of represented set of $g$ is a simple syntactic
abstraction not to be confused with the notion of set of values of
an expression $g$~\cite{AntoyHanus09PPDP}, which is a semantic
abstraction fairly more complicated.

\COMMENT{
\todo{NOTES:
If a choice has been pulled up, its two alternatives have
a non-empty fingerprint.
\\
If a choice is pulled above a node that has a non-empty
fingerprint, incompatibility can be detected locally.
}
}

\subsection{Invariance}

The proof of correctness of pull-tabbing
is based on two results that informally speaking 
establish that the notion of represented set
is invariant both by pull-tab steps and by non-choice steps.

\COMMENT{
\todo{Define $g$ and $g'$ are \emph{para-equal} iff
\emph{(1)}~ for any expression $e \in \rep_g$,
there exists an expression $e' \in \rep_{g'}$
such that $e = e'$ \modulo{}, and
\emph{(2)}~ for any expression $e' \in \rep_{g'}$, 
there exists an expression $e \in \rep_g$ 
such that $e = e'$ \modulo{}.
}
}

\def\lemmainvarpulltab{
\begin{lemma}[Invariance by pull-tab]
\label{invariance-by-pull-tab}
If $g \,\pulltab\, g'$ is a pull-tab step, then
\emph{(1)}~ for any expression $e \in \rep_g$,
there exists an expression $e' \in \rep_{g'}$
such that $e = e'$ \modulo{}, and
\emph{(2)}~ for any expression $e' \in \rep_{g'}$, 
there exists an expression $e \in \rep_g$ 
such that $e = e'$ \modulo{}.
\end{lemma}
\myproof{
We set up the notation for both claims.
If $g \,\pulltab\, g'$ is a pull-tab step,
then by Def.~\ref{def:pull-tab}:
\begin{displaymath}
  \begin{array}{@{} r c l @{}}
    g&=&C[n_f \col f(s_1,\ldots, n \col 
           ?(n_1 \col x, n_2 \col y),\ldots s_k)] \\[.75ex]
    g'&=&C[n' \col ?(n_{f_1} \col f (s_1,\ldots n_1 \col x,\ldots s_k),
                     n_{f_2} \col f(s_1,\ldots n_2 \col y,\ldots s_k))]
  \end{array}
\end{displaymath}
where $C$ is some context; $n_f$, $n$, $n_1$, $n_2$, $n'$, $n_{f_1}$ and $n_{f_2}$
are nodes; $f \ne\; ?$; $s_1,\ldots s_k$ and
$x$ and $y$ are expressions; 
$n$ is the $i$-th successor of $n_f$;
$n_1$, resp.{} $n_2$, is the $i$-th successor of $n_{f_1}$,
resp.{} $n_{f_2}$.
\\[1ex]
Claim (1):
let $A: g \toplus e$ be a computation
witnessing that $e \in \rep_g$, i.e., a consistent
computation making all and only the choice steps of $g$.
From this computation we construct a computation $A'$ of $g'$
that produces an expression $e'$ satisfying the claim.
Without loss of generality, since the notion of
represented set is well defined, we assume that the first 
step of $A$ is $g \to_{n,C_j} h$, for some $h$, 
where $C_j$ is either $C_1$ or $C_2$ of (\ref{binary-choice-rules}).
Let $\bar h$ and $h'$ be expressions defined by the following
computation: $g' \to_{n',C_j} \bar h \toequal_{n,C_j} h'$.
Node $n$ may or may not be in $g'$.
In particular, $n$ is in $g'$ iff $n$ has more
than one predecessor in $g$.  
If $n \in \nodes_{g'}$, then $n \in \nodes_{\bar h}$;
otherwise, the step $\bar h \toequal_{n,C_j} h'$ does not
replace any subexpression of $h'$ and $h'= \bar h$.
We explicitly construct a homomorphism $\rho : \nodes_h \to \nodes_{h'}$
that shows that $h = h'$ modulo a renaming of nodes.
We show that (a) for each node $m \in \nodes_h$,
with $m \ne n_{f_j}$, $m \in \nodes_{h'}$, and vice versa,
(b) for each node $m \in \nodes_{h'}$, with $m \ne n_f$,
$m \in \nodes_h$.
To prove (a), let $m \ne n_f$ be a node of $h$.
Since $g \to_{n,C_j} h$ is a choice step,
$m$ is either in the context $C$ of $g$
or in the subexpression rooted by $n_j$.
These portions of $g$ are preserved by the steps
that produce $g'$, $\bar h$ and $h'$.
Thus, $m \in \nodes_{h'}$.
The proof of (b) is analogous.
Therefore, we define $\rho(n_f)=n_{f_j}$
and $\rho(m)=m$, if $m \ne n_f$.
By (a) and (b) and by construction, $\rho$ is a bijection.
By construction, $\rho$ preserves root,
label, and successors of every node.
Thus the computation $A'$ starts with $g \to \bar h$,
followed by $\bar h \to h'$ if $\bar h \ne h'$.
Then, for any step of $A$ starting with expression $h$ at node
$p$ with rule $r$ there is a step of $A'$ starting with expression $h'$
at node $\rho(p)$ with rule $r$.
These computations start at equal expressions
(modulo a renaming of nodes)
and make the same steps, hence
they end at equal expressions (modulo a renaming of nodes).
Since $A$ is consistent, so is $A'$.
Let $e'$ be the last expression of $A'$.
This proves that $e' \in \rep_{g'}$.
\\[1ex]
Claim (2):
let $A': g' \toplus e'$ be a computation
witnessing that $e' \in \rep_{g'}$, i.e., a consistent
computation making all and only the choice steps of $g'$.
From this computation we construct a computation $A$ of $g$
that produces an expression $e'$ satisfying the claim.
Without loss of generality, since the notion of represented set
is well defined, we assume that $A'$ begins with the steps
$g' \to_{n',C_j} \bar h \toequal_{n,C_j} h'$.
The rule must be the same in both steps because,
if $n \in \nodes_{\bar h}$, then
$id(n)=id(n')$ and $A'$ is consistent.
Node $n$ may or may not be in $g'$.
In particular, $n$ is in $g'$ iff $n$ has more
than one predecessor in $g$.  
If $n \in \nodes_{g'}$, then $n \in \nodes_{\bar h}$;
otherwise, the step $\bar h \toequal_{n,C_j} h'$ does not
replace any subexpression of $h'$ and $h'= \bar h$.
We define the first step of $A$ as $g \to_{n,C_j} h$.
The rest of the proof is substantially equal to that of Claim (1).  
We complete $A$ with the same steps of $A'$ past $h'$
and obtain an expression $e$ in $\rep_{g}$.
We show in exactly the same way that,
modulo a renaming of nodes,
$h = h'$ and consequently $e = e'$.
}
}

\lemmainvarpulltab

\def\lemmainvarnonchoice{
\begin{lemma}[Invariance by non-choice]
\label{invariance-by-non-choice}
If $g \to g'$ is a rewrite non-choice step, then
\emph{(1)}~ for any expression $e \in \rep_g$,
there exists an expression $e' \in \rep_{g'}$
such that $e \tostar e'$ \modulo{}, and
\emph{(2)}~ for any expression $e' \in \rep_{g'}$, 
there exists an expression $e \in \rep_g$ 
such that $e \tostar e'$ \modulo{}.
\end{lemma}
\myproof{
Claim (1):
let $A:g=g_0 \to g_1 \to \ldots g_n=e$ be a computation
witnessing that $e \in \rep_g$, i.e., a consistent
computation making all and only the choice steps of $g$.
From $A$, we construct a computation $A'$ of $g'$
that produces an expression $e'$ satisfying the claim.
Consider the following diagram, where the top row is $A$
and the bottom row is $A'$:
\begin{displaymath}
    \xymatrix@C=12pt@R=16pt{
      \llap{$g =\;$} g_0 \ar[r] \ar[d] & g_1 \ar[r] \ar[d]^(.65)= 
          & \ldots & g_n \rlap{$\;=\;e$}  \ar[d]^(.65)= \\
      \llap{$g' = \;$} g'_0 \ar[r]^(.65)= & g'_1 \ar[r]^(.65)= 
          & \ldots & g'_n \rlap{$\; \tostar e'$}
    }
\end{displaymath}
By induction on $i$, for $i=1,\ldots n$, we both
define $g'_{i-1} \to g'_i$ 
and prove that the diagram commutes.
To support the induction, we strengthen the statement to
include the definition of the step $g_i \toequal g'_i$ and
the condition that this step is not a choice step.
Base case, $i=1$: The steps $g'_0 \leftarrow g_0 \to g_1$
are given by the assumptions.  Since the first is not a choice step
and the second is a choice step, they are at distinct nodes.
Hence, Lemma \ref{parallel-moves}
gives the steps $g'_0 \toequal g'_1 \leftequal g_1$ and
the commutativity of the diagram.
The step $g_1 \toequal g'_1$ is not a choice step because either
$g_1 = g'_1$ or it is at the same node as $g_0 \to g'_0$.
Ind.~case, $i>1$: The steps $g'_{i-1} \leftarrow g_{i-1} \to g_i$
are given by the induction hypothesis and assumption, respectively.
Since one is a choice step and the other is not, 
they are at distinct nodes. Hence,  Lemma \ref{parallel-moves}
gives the steps $g'_{i-1} \toequal g'_i \leftequal g_i$ and
the commutativity of the diagram.
The step $g_i \toequal g'_i$ is not a choice step because either
$g_i = g'_i$or it is at the same node as $g_{i-1} \to g'_{i-1}$.
Since $A$ is consistent, $A'$ up to $g'_n$ is consistent as well.
We reduce any remaining choice of $g'_n$ consistently with
the preceding steps of $A'$, say $g'_n \tostar e'$,
to produce an expression $e' \in \rep_{g'}$.
Thus, by the commutativity of the diagram
$e = g_n \toequal g'_n \tostar e'$ witnesses the claim.
\\[1ex]
Claim (2):
let $B: g' \tostar e'$ be a computation
witnessing that $e' \in \rep_{g'}$, i.e., a consistent
computation making all and only the choice steps of $g'$.
From this computation we construct a computation of $g$
that produces an expression $e$ satisfying the claim.
Let $Q = \nodes_g \cap \nodes_{g'}$, i.e., be the set of nodes
both in $g'$ and $g$.
Suppose that the cardinality of $Q$ is $n$, for some $n \geqslant 0$.
We reorder the steps of $B$, which is possible by
Lemma~\ref{parallel-moves}, so that any step at some node of $Q$
occurs before any step at some node not in $Q$.
Let $A':g'=g'_0 \to g'_1 \to \ldots~  g'_n \tostar e'$ 
be one such computation.
From $A'$, we construct a computation $A$ that produces
an expression $e$ satisfying the claim.
Consider the following diagram, where the top row is $A'$
and the bottom row is $A$:
\begin{displaymath}
    \xymatrix@C=12pt@R=16pt{
      \llap{$g' = \;$} g'_0 \ar[r] & g'_1 \ar[r]
          & \ldots & g'_n \ar[r]^(.65){*} & e' \\
      \llap{$g =\;$} g_0 \ar[r] \ar[u] & g_1 \ar[r] \ar[u]^(.65)= 
          & \ldots & g_n \ar[r]^(.65){*} \ar[u]^(.65)= 
          & e \ar[ul]_(.6){\kern-.65ex =} \\
    }
\end{displaymath}
By induction on $i$, for $i=1,\ldots n$, we both
define $g_{i-1} \to g_i$ and prove that the diagram commutes.
To support the induction, we strengthen the statement to
include the definition of the step $g_i \toequal g'_i$ and
the condition that this step is not a choice step.
Base case, $i=1$: 
Let $q_0$ be the root of the redex of $g'_0 \to_{q_0,r_0} g'_1$.
By assumption, $q_0 \in Q$. Hence $q \in \nodes_{g_0}$.
We let $g_0 \to_{q_0,r_0} g_1$.
Thus we have the steps $g'_0 \leftarrow g_0 \to g_1$
where by assumption the first is not a choice step and
by construction the second is a choice step.
Since these steps are at distinct nodes, by
Lemma \ref{parallel-moves} there exists some $g''$ such that
$g'_0 \toequal_{q_0,r_0} g'' \leftequal g_1$.
Therefore, $g''=g'_1$ and the diagram commutes.
The step $g_1 \toequal g'_1$ is not a choice step because either 
$g_1 = g'_1$ or it is at the same node as $g_0 \to g'_0$.
Ind.~case, $i>1$:
Let $q_{i-1}$ be the root of the redex of 
$g'_{i-1} \to_{q_{i-1},r_{i-1}} g'_i$.
By assumption, $q_{i-1} \in Q$, hence in $g$.
Thus, node $q_{i-1}$ in $g'_{i-1}$ is not created
by the step $g_{i-1} \toequal g'_{i-1}$.
Consequently, $q_{i-1}$ is a node of $g_{i-1}$ too.
We let $g_{i-1} \to_{q_{i-1},r_{i-1}} g_1$.
Thus we have the steps $g'_{i-1} \leftarrow g_{i-1} \to g_i$
where by assumption the first is not a choice step and
by construction the second is a choice step.
Since these steps are at distinct nodes, by
Lemma \ref{parallel-moves} there exists some $g''$ such that
$g'_{i-1} \toequal_{q_{i-1},r_{i-1}} g'' \leftequal g_i$.
Therefore, $g''=g'_i$ and the diagram commutes.
The step $g_i \toequal g'_i$ is not a choice step because either
$g_i = g'_i$ or it is at the same node as $g_{i-1} \to g'_{i-1}$.
Since $A'$ is consistent, $A$ up to $g_n$ is consistent as well,
since corresponding steps use the same rule.
We reduce any remaining choice of $g_n$ consistently with
the preceding steps of $A$, say $g_n \tostar e$,
to produce an expression $e \in \rep_{g}$.
We show that $e \toequal g'_n$.
If $e \ne g'_n$, then there exists
a choice step $g_n \to g_{n+1}$ in $A$.
By construction, this step is at some node $q_n$ which is not in $Q$
and hence is not in $g'_n$.
This means that the step $g_n \to g'_n$ erases node $q_n$.
Thus, we have the steps $g'_n \leftequal g_n \to_{q_n,r} g_{n+1}$,
for some rule  $r$, where by construction the first is not a choice step
and by assumption the second is a choice step. 
Since these steps are at distinct nodes, by
Lemma \ref{parallel-moves} there exists some $g''$ such that
$g'_n \toequal_{q_n,r} g'' \leftequal g_{n+1}$.
Since $q_n \not\in \nodes_{g'_n}$, $g''=g'_n$
and $g_{n+1} \toequal g'_n$.
The same above reasoning proves that for any expression $h$ of
$g_n \tostar e$ in $A$, $h \toequal g'_n$.
In particular, $e \toequal g'_n$.
Thus, $e \toequal g'_n \tostar e'$ witnesses the claim.
}
}

\lemmainvarnonchoice
We combine the previous lemmas
into computations of any length.

\def\corollsequence{
\begin{corollary}
\label{pre-correctness}
If $g \toxistar g'$ with no choice steps, then 
\emph{(1)}~ for any expression $e \in \rep_g$,
there exists an expression $e' \in \rep_{g'}$
such that $e \tostar e'$ \modulo{}, and
\emph{(2)}~ for any expression $e' \in \rep_{g'}$, 
there exists an expression $e \in \rep_g$ 
such that $e \tostar e'$ \modulo{}.
\end{corollary}
\myproof{
Both claims are proved by a trivial
induction on the number of steps of $g \toxistar g'$
using Lemmas~\ref{invariance-by-pull-tab} and
\ref{invariance-by-non-choice}.
}
}

\corollsequence

\def\theoremcorrect{
\begin{theorem}[Correctness]
\label{correctness}
If $g \toxistar g'$ with no choice steps, then 
\emph{(1)}~ for any value $v$ such that $g \tostar v$
is a consistent computation,
there exists a value $v'$ such that $g' \tostar v'$
is a consistent computation,
and $v = v'$ \modulo{}, and
\emph{(2)}~ for any value  $v'$ such that $g' \tostar v'$
is a consistent computation,
there exists a value $v$ such that $g \tostar v$
is a consistent computation,
and $v = v'$ \modulo{}.
\end{theorem}
\myproof{
Claim (1):
let $A : g \tostar v$ a consistent computation of $g$
into a value $v$.
By Lemma~\ref{parallel-moves}, without loss of generality
we assume that $A : g \tostar e \tostar v$,
where the segment $g \tostar e$ consists of all
the choice steps of $g$.
Since $A$ is consistent, $e \in \rep_g$.
By Corollary~\ref{pre-correctness}, there exists
a consistent computation $g' \tostar e'$ such that 
$e = e'$ (modulo a renaming of nodes).
Since $e = e'$ (modulo a renaming of nodes) and $e \tostar v$,
there exists a computation $e' \tostar v'$
such that $v = v'$ (modulo a renaming of nodes).
\\[1ex]
Clam (2): the proof is analogous to that of claim (1).
}
}

\theoremcorrect
Theorem \ref{correctness} suggests to apply both non-choice and
pull-tab steps to an expression. Choices pulled up to the root are
reduced consistently and without context cloning. Of course, by
the time a choice is reduced, all its spines have been
cloned---similar to bubbling and copying.
A better option, available to pull-tabbing only,
is discussed in the next section.

\section{Application}
\label{Application}

The pull-tab transformation is meant to be used in conjunction
with some evaluation strategy.  
We showed that pull-tabbing is not tied to any particular strategy.
However, the strategy should be pull-tab-aware in that:
(1) a choice should be evaluated (to a head normal form)
only when it is \emph{needed} \cite{Antoy97ALP},
(2) a choice in a root position is reduced (consistently),
whereas in a non-root position is pulled, and
(3) before pulling a choice,
one of the choice's alternatives should be a head-normal form.
The formalization of such a strategy
would take us well beyond the scope of this paper.

In well-designed, non-deterministic programs,
either or both alternatives of most (but not all) choices
should fail \cite{Antoy10JSC}.
%
Under the assumption that a choice is evaluated to a head normal form
only when it is \emph{needed} \cite{Antoy97ALP},
if an alternative of the choice fails,
the choice is no longer non-deterministic---the
failing alternative cannot produce a value.
Thus, the choice can be reduced to the other alternative
without loss of completeness and without context cloning.
This is where pull-tabbing is advantageous over
copying and bubbling---any portion of a choice's context not yet cloned
when an alternative fails no longer needs to be cloned.
Of course, the implementation must still identify the choice,
and choice's single remaining strand as either left or right, 
to ensure consistency.

\section{Related Work}

We investigated pull-tabbing, an approach to
non-deterministic computations in functional logic programming.
Section \ref{Approaches} recalls copying and bubbling,
the competitors of pull-tabbing.
Here, we briefly highlight the key differences between these approaches.
Pull-tabbing ensures the completeness of computations
in the sense that no alternative of a choice is left behind
until all the results of some other alternative have been produced.
Similar to every approach with this property, it must clone
portions of the context of a choice.
In contrast to copying and bubbling, it clones
the context of a choice in minimal increments
with the intent and the possibility of stopping cloning
the context as soon as an alternative of the choice fails.

The idea of identifying choices to avoid combining in some
expression the left and right alternatives of the same choice
appears in \cite{BrasselHuchAPLAS07}. The idea is developed
in the framework of a natural semantics for the translation
of (flat) Curry programs into Haskell.
A proof of the correctness of this idea will appear in
\cite{Brassel2011PhD} which also addresses the similarities
between the natural semantics and graph rewriting.  
This discussion, although informal, is enlightening.

\section{Conclusion}

We formally defined the pull-tab transformation,
characterized the class of programs for which the transformation
is intended, extended the computations in these programs
to include the transformation, proved the correctness
of these extended computations, and described the condition
that reduces context cloning.
In contrast to its competitors, in pull-tabbing any step is a
simple and localized graph transformation.
This fact should ease executing the steps in parallel.
Future work, aims at defining a pull-tab-aware parallel strategy
and implementing it to measure the effectiveness of pull-tabbing.

\newpage
\bibliographystyle{acmtrans}

\end{document}